\documentclass[preprint]{aastex}
\shorttitle{Stability of Toomre-Hayashi Disk}
\shortauthors{Hanawa, Saigo \& Matsumoto}
\begin{document}

\title{Stability of Toomre-Hayashi Disk}

\author{Tomoyuki Hanawa and Kazuya Saigo}
\affil{Department of Astrophysics, School of Science, Nagoya University,
Chikusa-ku, Nagoya 464-8602, Japan}
\email{hanawa@a.phys.nagoya-u.ac.jp, saigo@a.phys.nagoya-u.ac.jp}

\and

\author{Tomoaki Matsumoto} 
\affil{Department of Humanity and Environment, Hosei University,
Fujimi, Chiyoda-ku, Tokyo 102-8160, Japan}
\email{matsu@i.hosei.ac.jp}

\begin{abstract}
We investigate stability of Toomre-Hayashi model for a
self-gravitating, rotating gas disk.
The rotation velocity, $ v _\varphi $, and sound speed,
$ c _{\rm s} $, are spatially constant in the model.
We show that the model is unstable against an axisymmetric
perturbation irrespectively of the values of 
$ v _\varphi $ and $ c _{\rm s} $.
When the ratio, $ v _\varphi / c _{\rm s} $, is large,
the model disk is geometrically thin and unstable against
an axisymmetric perturbation having a short radial wavelength,
i.e., unstable against fragmentation.
When $ v _\varphi / c _{\rm s} $ is smaller than 1.20, it is 
unstable against total collapse.
Toomre-Hayashi model of 
$ 1.7 \, \la \, v _\varphi / c _{\rm s} \, \la \, 3 $ 
was thought to be stable against axisymmetric perturbations
in earlier studies in which only radial motion was taken into
account.  Thus, the instability of Toomre-Hayashi model
having a medium $ v _\varphi / c _{\rm s}  $ is due to not change in the
surface density but that in the disk height.
We also find that the singular isothermal
perturbation is stable against non-spherical peturbations.
\end{abstract}

\keywords{gravitation --- 
hydrodynamics --- instabilities --- stars: formation}

\section{INTRODUCTION}

\citet{toomre82} and \citet*{hayashi82}
found  an analytic model for a self-gravitating gas cloud.
The model describes a disk in which the self-gravity
balances with the centrifugal force and gas pressure.
The rotation velocity, $ v _\varphi $, and the sound
speed, $ c _{\rm s} $, are assumed to be constant.
The density distribution of the model is expressed as
\begin{equation}
\rho \, (r, \, \theta) \; = \; \frac{\gamma ^2 c _{\rm s} ^2}
{2 \pi G r ^2} \, 
\left\lbrack \frac{\cosh \zeta}{\cosh (\gamma \zeta)} 
\right\rbrack ^2 \; ,
\label{th-model}
\end{equation}
where
\begin{equation}
\zeta \; = \; \ln \, \left(
\frac{1 \, + \, \cos \, \theta}{\sin \, \theta} \right) \; ,
\end{equation}
and
\begin{equation}
\gamma \; = \; 1 \, + \, \frac{v _\varphi ^2}{2 c _{\rm s} ^2} 
\; ,
\end{equation}
in the spherical polar coordinates, ($ r $, $ \theta $,
$ \varphi $).

This simple model is useful for studying dynamics of
self-gravitating gas disk.  Earlier researchers
\citep*[see, e.g.,][]{kalnajs71, lemos91, syer96,
evans98a, evans98b, goodman99, shu00} studied 
further simplified model, i.e., the razor-thin power law disk
model, which is equivalent to the vertically averaged 
Toomre-Hayashi model. 
Besides of its simplicity, it is a good approximation to
a centrifugally supported disk formed by collapse of
a rotating gas cloud \citep*[see, e.g.,][]
{matsumoto97, saigo98}.
The centrifugally supported disk will evolve into 
single or binary stars associated with gas disks.
The evolution may be driven by instabilities of the disk.
Since the Toomre-Hayashi model is an exact equilibrium
solution, it is ideal for linear stability analysis;
we need not worry about inaccuracy of the unperturbed state.

In this paper we concentrate on the simplest mode, the
axisymmetric one.  We will show that the Toomre-Hayashi
model is unstable for any $ \gamma $, i.e., any
$ v _\varphi / c _{\rm s} $.  This result is different from
prediction by \citet{hayashi82} and  earlier analyses by 
\citet{lemos91} and \citet{shu00}.  While 
two-dimensional perturbation is taken into account in our
analysis, only radial perturbation was considered in
earlier analysis.  We believe that our analysis will provide
insights for stability of self-gravitating disk.

The rest of our paper has the following structure.  In \S 2
we derive energy principle for axisymmetric perturbation
around the Toomre-Hayashi disk.  In \S 3 we obtain the
marginally stable mode for the Toomre-Hayashi disk. 
In \S 4 we show that it is unstable for any $ \gamma $ and
discuss implications of our stability analysis.
Main conclusions are summarized in \S 5.

\section{PERTURBATION EQUATION AND ENERGY PRINCIPLE}

We consider a small perturbation around the Toomre-Hayashi
disk in the polar coordinates.
In this paper we use the displacement vector,
$ \mbox{\boldmath$\xi$} \, = \, (\xi _r, \, \xi _\theta ) $, 
in the $ \varphi \, = \, {\rm const.} $ plane to denote a perturbation.
From the equation of continuity we can evaluate the change 
in the density as
\begin{equation}
\rho _1 \; = \; - \, \frac{1}{r ^2} \,
\frac{\partial}{\partial r} \, (r ^2 \rho _0 \xi _r) \;
- \, \frac{1}{r \sin \theta} \, \frac{\partial}{\partial \theta}
\, (\sin \theta \rho _0 \xi _\theta ) \; ,
\label{continuity}
\end{equation}
where $ \rho _0 $ denotes the density in the equilibrium
and is given by Equation (\ref{th-model}).
From the angular momentum conservation we can evaluate
the change in the rotation velocity as
\begin{equation}
v _{\varphi 1} \; = \; - \, \frac{v _{\varphi _0}}{r} \,
\left( \xi _r \, + \, \frac{\xi _\theta}{\tan \, \theta}
\right) \; , \label{amomentum}
\end{equation}
where $ v _{\varphi 0} $ denotes the rotation velocity in
equilibrium.  The perturbation equation for the 
equation of motion is expressed as
\begin{equation}
\frac{\partial ^2 \xi _r}{\partial t^2} \; - \;
\frac{2 \, v _{\varphi 0} v _{\varphi 1}}{r} \; + \;
c _{\rm s} ^2 \, \frac{\partial}{\partial r} \,
\left( \frac{\rho _1}{\rho _0} \right) \; + \;
\frac{\partial \phi _1}{\partial r} \; = \; 0 \; ,
\label{motion-r}
\end{equation}
and
\begin{equation}
\frac{\partial ^2 \xi _\theta}{\partial t^2} \;
\; - \; \frac{2 \, v _{\varphi} v _{\varphi 1}}{r \, \tan \theta}
+ \; \frac{c _{\rm s} ^2}{r} \, \frac{\partial}{\partial \theta}
\, \left( \frac{\rho _1}{\rho _0} \right)  \; + \;
\frac{1}{r} \, \frac{\partial \phi _1}{\partial \theta} \;
\; = \; 0 \; ,
\label{motion-theta}
\end{equation}
for the $ r $- and $ \theta $-directions, respectively.
The change in the gravitational potential, $ \phi _1 $,
is related to that in the density by
\begin{equation}
\left\lbrack \frac{1}{r ^2} \, \frac{\partial}{\partial r}
\, \left( r ^2 \, \frac{\partial}{\partial r} \right)
\; + \; \frac{1}{r ^2 \sin \theta} \,
\frac{\partial}{\partial \theta} \,
\left( \sin \theta \, \frac{\partial}{\partial \theta}
\right) \, \right\rbrack \,
\phi _1 \; = \; 4 \pi G \, \rho _1 .
\label{poisson}
\end{equation}

Before detailed manipulation we consider the characteristic
of the Tooomre-Hayashi model.  Since the Toomre-Hayashi model
has neither inner nor outer boundary, its global analysis
is not straight forward \citep[see]
[for mathematical discussions on the stability
of scale-free disks]{goodman99}.

The Tooomre-Hayashi disk  is scale free and has no 
characteristic length scale.  It has neither 
characteristic timescale; both the rotation period and
sound crossing time are proportional to the radial distance, 
$ r $. Since the disk has no
characteristic radius, inner and outer parts of a disk are
equally either stable or unstable.  This means that the disk
has no discrete eigenmode which grows exponentially in time
\citep[see also][]{goodman99}.

If the Toomre-Hayashi disk is unstable, the perturbation grows
faster in an inner part and propagate outwards since the dynamical
timescale is shorter at a smaller radius.
To show this  we derive the energy principle,
\begin{equation}
{\cal T} \; + \; {\cal W} \; + \; {\cal S} \; = \; 0 \;
\label{energy}
\end{equation}
where
\begin{eqnarray}
{\cal T} & = & \int _a ^b \int _0 ^\pi \rho _0  \left\vert
\xi _r ^* \frac{\partial ^2 \xi _r}{\partial t^2}
\, + \, \xi _\theta ^* \frac{\partial ^2 \xi _\theta}{\partial t ^2}
\right\vert \, r ^2 \sin \theta \;
dr d\theta \; , \\
{\cal W} & = & 
\int _a ^b \int _0 ^\pi \left\{ \rho _0 \left\lbrack
2 \, \vert v _{\varphi 1} \vert ^2
\, + \, c _{\rm s} ^2 \,
\left\vert \frac{\rho _1}{\rho _0} \right\vert ^2
\right\rbrack \, + \, \rho _1 ^* \phi _1 \right\} \,
r ^2 \sin \theta \, dr d\theta \; , \\
{\cal S} & = & \left\lbrack
\int _0 ^\pi r ^2 \, \sin \theta \,
\xi _r ^* \, ( c _{\rm s} ^2 \rho _1
\, + \, \rho _0 \phi _1 ) \, d\theta \right\rbrack _a ^b 
\; . \label{surface} 
\end{eqnarray}
Equation (\ref{energy}) is derived from equations
(\ref{continuity}) through (\ref{poisson}) by multiplying 
the complex conjugate of $ \rho _1 $, $ \xi _r $, $ \xi _\theta $,
and $ \phi _1 $.  We can set the inner and outer radii, 
$ a $ and $ b $, arbitrarily. 
Equation (\ref{energy}) denotes the
energy conservation in the second order in the interval of
$ a \, \le \, r \, \le \, b $.

The term, $ {\cal T} $,
denotes the product of acceleration and displacement.
If it is positive, the perturbation grows and the system is
unstable.  The term, $ {\cal W} $, denotes the excess of rotation
energy, thermal energy and gravitational energy. 
We consider the condition that the term, $ {\cal S} $, vanishes.
This condition is equivalent to 
\begin{eqnarray}
\xi _r (r, \, \theta) & = & \xi _r (\theta) \, r ^{ik+1/2} \; ,
\label{power1} \\
\xi _\theta (r, \, \theta) & = & \xi _\theta (\theta) \, r ^{ik+1/2} \; , \\
\rho _1 (r, \, \theta) & = & \rho _1 (\theta) \, r ^{ik-5/2} \; ,\\
\phi _1 (r, \, \theta) & = & \phi _1 (\theta) \, r ^{ik-1/2} \; , 
\label{power4} \\
\end{eqnarray}
where $ k $ denotes the wavenumber in the logarithmic scale
(=~$ \partial /\partial \ln r $).  When $ k \, \ln (b/a) \, = \,
2 \pi $, the third integral vanishes in equation (\ref{energy}).
We call this the periodic boundary condition in the following.
Our periodic boundary condition is equivalent to the surface
density perturbation assumed by \citet{kalnajs71} and others.

When we apply the periodic boundary condition, the term
$ {\cal S} $ is proportional to $ \ln \, (b/a) $ and scale free.
Also the term $ {\cal T} $ should be scale free and the
growth timescale should be proportional to the radius,
$ \partial /\partial t \, \propto \, 1 / r $.  This means
that there exists no exponentially growing mode satisfying the
periodic boundary condition.  We seek, instead,  
a marginally stable mode satisfying the periodic boundary condition.

Substituting the boundary condition into the perturbation 
equation we obtain
\begin{equation}
\left( i k \, + \, \frac{1}{2} \right) \, \xi _r \; + \;
\frac{1}{\rho _0} \, \frac{\partial \rho _0}{\partial \theta} \, \xi
_\theta
\; + \; \frac{1}{\sin \theta} \, \frac{\partial}{\partial \theta}
\, (\sin \theta \, \xi _\theta) \; + \;
\left( \frac{\rho _1}{\rho _0} \right) \; = \; 0
\; ,
\label{mass}
\end{equation}
\begin{equation}
2 v _{\varphi 0} v _{\varphi 1}
\, + \, \left( i k \, - \, \frac{1}{2} \right) \,
c _{\rm s} ^2 \, \left( \frac{\rho _1}{\rho _0} \right) \; + \;
\left( i k \, - \, \frac{1}{2} \right) \, \phi _1 \; = \; 0
\; ,
\label{motionr}
\end{equation}
\begin{equation}
- \, \frac{2 v _{\varphi 0} v _{\varphi 1}}{\tan \theta} \; + \;
c _{\rm s} ^2 \, \frac{\partial}{\partial \theta} \,
\left( \frac{\rho _1}{\rho _0} \right) \; + \;
\frac{\partial \phi _1}{\partial \theta} \; = \; 0
\; ,
\label{motionp}
\end{equation}
\begin{equation}
v _{\varphi 1} \; = \; - \, v _{\varphi 0}
\, \left( \xi _r \, + \, \frac{\xi _\theta}{\tan \theta}
\right) \; ,
\end{equation}
\begin{equation}
\left\lbrack - \, \left( k ^2 \, + \, \frac{1}{4} \right)
\; + \;
\frac{1}{\sin \theta} \, \frac{\partial}{\partial \theta}
\, \left( \sin \theta \, \frac{\partial}{\partial \theta}
\right) \right\rbrack \, \phi _1 \; = \; 4 \pi G \rho _1
\; , \label{poisson-t}
\end{equation}
where
\begin{equation}
\rho _0 \; = \; \frac{\gamma ^2 c _{\rm s} ^2}{2 \pi G} \,
\left\lbrack \frac{\cosh \, \zeta}{\cosh \, (\gamma \zeta)}
\right\rbrack ^2
\end{equation}
and
\begin{equation}
v _{\varphi 0} \; = \; c _{\rm s} \, 
\sqrt{2 \, (\gamma \, - \, 1)} \, .
\end{equation}
We solve equations (\ref{mass}) through (\ref{poisson-t})
as an eigenvalue problem for $ k $.

\section{MARGINALLY STABLE MODE}

\subsection{Singular Isothermal Sphere}

When $ \gamma \, = \, 1 $, the Toomre-Hayashi disk 
is spherically symmetric and reduces to the singular 
isothermal sphere.  The density distribution is expressed as
\begin{equation}
\rho _0 \; = \; \frac{c _{\rm s} ^2}{2 \pi G r ^2} \; .
\end{equation}
Then the perturbation should be
expressed by the spherical harmonics, 
$ Y _\ell ^m (\theta, \, \varphi) $.  We concentrate 
on the axisymmetric perturbation for simplicity. Then
the density perturbation is expressed as
\begin{equation}
\rho _1 (\theta) \; = \; A \, P _\ell (\cos \, \theta) \; ,
\end{equation}
where $ A $ and $ P _\ell $ denotes an arbitrary scale
factor and the Legendre's polynomial, respectively.
Then the change in the potential is given by
\begin{equation}
\phi _1 \; = \; - \, 4 \pi G \rho _1 \,\left\lbrack k ^2 \,
+ \, \frac{1}{4} \, + \, 
\ell \, (\ell \, + \, 1) \right\rbrack ^{-1} 
\; . \label{poisson-s}
\end{equation}
Substituting equation (\ref{poisson-s}) into equation
(\ref{motionp}) we obtain
\begin{equation}
k ^2 \; = \; \frac{7}{4} \, - \, 
\ell \, (\ell \, + \, 1) \; .
\label{singular}
\end{equation}
From equation (\ref{singular}) we conclude that
the marginally stable mode exists only for $ \ell \, = \, 0 $
and the critical wavenumber is $ k \, = \, \sqrt{7}/2 $.

The critical wavenumber coincides with that of the
equilibrium isothermal sphere.  As noted in
the textbook of \citet{chandra39},
the density distribution is approximately given by
\begin{equation}
\rho \; = \; \frac{c _{\rm s} ^2}{2 \pi G r ^2} \,
\exp \, \left\{ \frac{A}{\sqrt{r}}
\, \cos \, \left\lbrack \frac{\sqrt{7}}{2} \,
\log \, \left( \frac{r \sqrt{4 \pi G \rho _{\rm c}}}{c _{\rm c}} \right) 
\, - \, \delta \right\rbrack 
\right\} \; .
\label{chandra}
\end{equation}
in the region of $ r \, \gg \, c _{\rm s} / \sqrt{4 \pi G \rho _{\rm c}} $,
where $ \rho _{\rm c} $ denotes the central density.
This equilibrium isothermal sphere has almost the same density 
distribution as that of the singular isothermal sphere in
the region of $ r \, \gg \, c _{\rm s} / \sqrt{4 \pi G \rho _{\rm c}} $.
The small difference between them can be regarded as
a neutral perturbation on the singular isothermal sphere.
Note the wavenumber, $ \sqrt{7} /2 $, in Equation (\ref{chandra}). 

The singular isothermal sphere is stable
against non-spherical perturbation of $ \ell \, \ne \, 0 $.
Equation (\ref{singular}) leaves formally a possibility that
the singular isothermal disk is unstable for any $ k $. 
This possibility is, however, very unlikely.  
When $ k $ is very large, the wavelength is so short to
destabilizes the gas sphere.  Thus we can exclude the possibility
that the singular isothermal sphere is unstable.

\subsection{Toomre-Hayashi Disk}

In this subsection we solve the perturbation equation
to obtain the critical wavenumber $ k $ as a function of 
$ \gamma $.  First we obtain
\begin{equation}
\frac{\partial v _{\varphi 1}}{\partial \theta} \, + \, 
\tan \theta \, \left(ik \, - \, \frac{1}{2} \right) \, 
v _{\varphi 1} \; = \; 0 \; ,
\label{vphi}
\end{equation}
from equations (\ref{motionr}) and (\ref{motionp}).
Equation (\ref{vphi}) has the solution,
\begin{equation}
v _{\varphi 1} \; \propto \; (\sin \, \theta) ^{ik \, - \, 1/2} \; .
\end{equation}
To avoid the divergence of $ v _{\varphi 1} $ at $ \theta \, = \, 0 $, 
we seek modes of $ v _{\varphi 1} \, = \, 0 $.

Substituting $ v _{\varphi 1} \, = \, 0 $ into
equations (\ref{mass}) through (\ref{poisson-t}), we obtain
\begin{equation}
\xi _r \; = \; - \, \frac{\xi _\theta}{\tan \, \theta} \; 
\label{displacement2}
\end{equation}
\begin{equation}
c _{\rm s} ^2 \, \frac{\rho _1}{\rho _0} \; = \; - \,
\phi _1 \; ,
\end{equation}
\begin{equation}
- \, \left( i k \, + \, \frac{1}{2} \right) \,
\frac{\xi _\theta}{\tan \theta} \; + \;
\frac{1}{\rho _0} \, \frac{\partial \rho _0}{\partial \theta} \, \xi
_\theta
\; + \; \frac{1}{\sin \theta} \, \frac{\partial}{\partial \theta}
\, (\sin \theta \, \xi _\theta) \; - \;
\left( \frac{\phi _1}{c _{\rm s} ^2} \right) \; = \; 0 \; ,
\label{mass2}
\end{equation}
and
\begin{equation}
\left\lbrack - \, \left( k ^2 \, + \, \frac{1}{4} \right)
\; + \;
\frac{1}{\sin \theta} \, \frac{\partial}{\partial \theta}
\, \left( \sin \theta \, \frac{\partial}{\partial \theta}
\right) \right\rbrack \, \phi _1 \; = \; - \,
\frac{4 \pi G \rho _0}{c _{\rm s} ^2} \, \phi _1 \; .
\label{poisson-t2}
\end{equation}

We solve Equation (\ref{poisson-t2}) with the boundary
conditions $ \partial \phi _1 / \partial \theta \, = \, 0 $
at $ \theta $ = 0 and $ \pi $ as an eigenvalue problem.
Once $ \phi _1 $ is given, other perturbations, 
$ \xi _r $, $ \xi _\theta $, and $ \rho _1 $ are derived 
from Equations (\ref{displacement2}) through (\ref{mass2}).
For a given set of $ \gamma $ and $ k $, 
we integrated Equation (\ref{poisson-t2}) with the Runge-Kutta 
method.  By try and error, we found pairs of $\gamma $ and $ k $ 
for which Equation (\ref{poisson-t2}) has a solution satisfying
the boundary condition at $\theta $ =  0 and $ \pi $.

Figure 1 denotes the eigenvalue, $ k $, as a function of $ \gamma $.
The solid  curve connects the crosses which denotes the
numerical obtained eigen values.
The dashed curve does $ k \, = \, \gamma $ for comparison.
It is a good approximation to the numerically obtained 
eigenvalues for $ \gamma \, \gg \, 1 $.

Figure 2 shows $ \rho _0/ \rho _1 $ as a function of $ \theta $ for
the marginally stable modes.  Each curve is labeled with $ \gamma $.
The relative density perturbation, $ \rho _1 / \rho _0 $,
is symmetric with respect to $ \theta \, = \, \pi / 2 $ and
has no zero point.
When $ \gamma $ is very large, the density perturbation is
restricted near $ \theta \, = \, \pi / 2 $.  Note that Figure 2
shows not $ \rho _1 $ but $ \rho _1 / \rho _0 $.
The density perturbation, $ \rho _1 $, is very small near the
pole ($ \theta \, = \, 0 $ and $ \pi $) even for 
$ \gamma \, = \, 2 $, since $ \rho _0 $ is very small there. 

After having completing numerical computations, we found an
asymptotic solution for Equation (\ref{poisson-t2}).
To derive the asymptotic solution, we rewrite Equation 
(\ref{poisson-t2}) into
\begin{equation}
\left\lbrack - \, \left( k ^2 \, + \, \frac{1}{4} \right)
\; + \; \cosh ^2 \zeta \,
\frac{\partial ^2}{\partial \zeta ^2} \,
\right\rbrack \, \phi _1 \; = \; - \,
2 \gamma ^2 \, \left\lbrack \frac{\cosh \zeta}{\cosh (\gamma \zeta)}
\right\rbrack ^2 \, \phi _1 \; .
\label{poisson-zeta}
\end{equation}
When $ \gamma \, \gg \, 1 $, $ \phi _1 $ has an appreciable
amplitude only near the disk plane 
(i.e., $ \theta \, \simeq \, \pi / 2 $ or equivalently
$ \zeta \, \simeq \, 0 $).  Thus, Equation
(\ref{poisson-zeta}) can be approximated as
\begin{equation}
\left\lbrack - \, \left( k ^2 \, + \, \frac{1}{4} \right)
\; + \; 
\frac{\partial ^2}{\partial \zeta ^2} \,
\right\rbrack \, \phi _1 \; = \; - \,
\frac{2 \gamma ^2}{\cosh ^2 (\gamma \zeta)}
\, \phi _1 \; .
\label{poisson-zeta2}
\end{equation}
Equation (\ref{poisson-zeta2}) has an analytic solution
\begin{equation}
\phi _1 \; = \; \frac{1}{\cosh \, (\gamma \zeta)} \; ,
\label{asymptotic}
\end{equation}
for $ k ^2 \, = \, \gamma ^2 \, - \, 1/4 $.
This denotes an asymptotic solution in the limit of
$ \gamma \, \gg \, 1 $.

Figures 3 and 4 show the marginally stable perturbation by cross section.
Figure 3 shows the density distribution in the Toomre-Hayashi
model of $ \gamma \, = \, 10 $.  The solid curves are
the contours of $ \log \, \rho \, = \, n/4 $ where $ n $ is
an integer.  Only the sector near the disk plane is shown in
Figure 3.  Figure 4 is the same as
Figure 3 but the marginally stable mode is superimposed on the
Toomre-Hayashi model.  As shown in Figure 4 the neutrally
stable mode deforms the Toomre-Hayashi model into rings.

\section{DISCUSSION}

\subsection{Instability Strip}

We have shown in the previous section that the Toomre-Hayashi 
model has a marginally stable mode for any $ \gamma $. 
This means that the Toomre-Hayashi model of any $ \gamma $ 
is unstable against axisymmetric perturbation, since
a marginally stable mode is adjacent to an unstable mode. 

As shown by \citet{hayashi82}, the Toomre-Hayashi model is
unstable for total collapse when $ \gamma \, < \, 2.441 $.
In other words, the Toomre-Hayashi model of $ \gamma \, = \, 2.441 $
has a  marginally stable mode of $ k $ = 0.  This marginally stable
point is also shown by an asterisk in Figure 1.
It is also shown in the previous section that the Toomre-Hayashi model of
$ \gamma \, = \, 0 $ (i.e., the singular isothermal sphere) is
unstable when $ k \, < \, \sqrt{7}/2 $.  Accordingly, we conclude
that the Toomre-Hayashi model is unstable in the lower right side
of the marginally stable curve in Figure 1. 

The above conclusion apparently contradicts with the 
stability analyses by \citet{lemos91}
and \citet{shu00} and prospect by
\citet{hayashi82}.   All of them  concluded that
the Toomre-Hayashi model of $ 2.5 \, \la \, \gamma \, \la \, 6 $ 
is stable against axisymmetric perturbation.

The dash-dotted curves of Figure 1 denote the points of marginally stability
obtained by \citet{shu00}.  They analyzed the stability of
the Toomre-Hayashi disk under the thin disk approximation while
taking account of magnetic fields.  The magnetic field reduces
the effective gravity and increases the sound speed 
\citep[see, e.g.,][]{li97, shu97, nakamura97}.
The effects can be
fully taken into account by replacing the gravitational constant
($ G $) and sound speed ($ c _{\rm s} $) with the effective ones.
Thus the stability of the magnetized disk is essentially 
the same as that of
non-magnetized one.  The condition for the marginal stability
of the axisymmetric mode is denoted by
\begin{equation}
\gamma \; = \; 1 \, + \, 
\frac{\left(k ^2 \, + \, 1/4 \right)
\, \lbrack K (k) \, - \, 1 \rbrack}
{2 \, - \, \left( k ^2 \, + \, 1/4 \right) \, K (k)}  
\; , \label{radial}
\end{equation}
where
\begin{equation}
K (k) \; = \; \frac{1}{2} \,
\frac{ \Gamma \, \lbrack (1/2 \, + \, i k) /2 \rbrack \,
\Gamma \, \lbrack (1/2 \, - \, i k) /2 \rbrack}
{ \Gamma \, \lbrack (3/2 \, + \, i k) /2 \rbrack \,
\Gamma \, \lbrack (3/2 \, - \, i k) /2 \rbrack} \; ,
\end{equation}
denotes the Kalnajs function for $ m \, = \, 0 $.
\citet{lemos91} obtained also the same dispersion relation
although Equation (\ref{radial}) was not explicitly printed. 

As shown in Figure 1,  
the disturbance is oscillatory and stable when $ \gamma $
is medium.  
This discrepancy is due to incompleteness in their search for
unstable modes.  \citet{shu00} considered only radial
perturbations neglecting vertical structure.  However, 
vertical perturbation is dominant in the marginally stable mode 
that we found in this paper.  Thus they could not find 
this marginally stable mode and adjacent unstable mode.

Equation (\ref{radial}) has an asymptotic form of
\begin{equation}
\gamma \, = \, k \, + \, 2 \, + \, \frac{15}{8 k} \,
+ \, {\cal O} \, ( k ^{-2}) \; ,
\end{equation}
in the limit of $ k \, \gg \, 3 $.  Note that the solid and
dash-dotted curves are parallel in the region of large $ k $
and hence for large $ \gamma $.
Change in the vertical structure has a significant effect
even for a disk of a large $ \gamma $.
Comparison of our analysis and that of Shu et al. (2000) tells
us importance of vertical perturbation in the instability of
the Toomre-Hayashi model.  The importance will be also true
for stability of self-gravitating disk in general.

\citet{hayashi82} prospected on the basis of
their numerical simulations that the Toomre-Hayashi is
stable when $ \gamma $ is medium.
Their search  was also incomplete since the
the dynamic range in the radial direction is restricted.
They could not find an unstable mode having a small $ k $.
Note that the wavenumber of the marginally stable mode is
3.102 and 5.058 for $ \gamma $ = 3 and 5, respectively.
This means that the radius increases by a factor of
7.58 and 3.46, respectively, at every node of the marginally 
stable mode.  Thus large dynamic range is necessary to 
find an unstable mode from numerical simulations.
We think that we need dynamic range covering at least two
wavelengths.

As shown in the previous section, the change in the rotation
velocity ($ \delta v _\varphi $) vanishes in the marginally
stable mode.  This means that the displacement is only in
the vertical direction.  The marginally stable mode
changes hight of the disk but not the surface density.

\subsection{Eigenmode}

Although we have tried, we could not find the lower limit
on $ k $ for a given $ \gamma $.  In other words, we failed
in finding the other side of the instability strip.  This
failure is not due to lack in our numerical technic.

One might think that the other marginally stable mode would
stem from the edge of the total collapse, 
$ (k, \, \gamma) \, = \, (0, \, 0.2441) $. 
It, however, cannot stem.  The marginally stable total
collapse has radial dependence different from those assumed in
Equations (\ref{power1}) through (\ref{power4}).  
The change in the density distribution is
expressed as
\begin{eqnarray}
\rho _1 & = & \frac{\partial \rho _0}{\partial \gamma} \\
& = & \left\lbrack \frac{2}{\gamma} \, - \, 2 \zeta \, 
\tanh \, (\gamma \zeta) \right\rbrack \, \rho _0 \\
& \propto & r ^{-2} \, ,
\end{eqnarray}
for the marginally stable total collapse.  This total collapse
was called the breathing mode in \citet{lemos91}.

The other marginally stable mode might have different radial
dependence which have been not yet known.  
Usually we assume some temporal and spatial dependences 
when we search unstable modes.  Often we assume that the
perturbation grows exponentially.  An eigenmode may not
grow exponentially in time in the Toomre-Hayashi disk.
Remember our argument based on the energy principle in
\S 2 
\citep[see also][for another discussion
on the stability of scale-free disks]{goodman99}.

The expansion wave solution of \citet{shu77} gives us a hint.
The eigenmode may have a form of
\begin{equation}
\rho _1 \; = \; r ^{-2} \,  
f \left( \frac{r}{c _{\rm s} t}, \, \theta \right) 
\end{equation}
The density distribution is expressed as
\begin{equation}
\rho \; = \; \frac{c _{\rm s} ^2}{2 \pi G r ^2} \,
g \left( \frac{r}{c _{\rm s} t} \right) \; ,
\end{equation}
in the expansion wave solution.  The difference from
the singular isothermal sphere can be regarded as a 
perturbation propageting outwards while growing.
As pointed out by \citet{hunter77}, the singular isothermal
sphere can be regarded as a similarity solution. 
Similarly the Toomre-Hayashi model can be regarded as
a similarity solution.  An unstable mode can grow in proportion
to the power of time when the unperturbed state is a similarity
solution \citep[see, e.g.,][]{hanawa97}.

\section{CONCLUSION}

We discussed the stability of the Toomre-Hayashi model against
axisymmetric perturbations in this paper.  Our approach was 
conservative since we focused on the marginally stable
mode.  We did not obtain an unstable mode.  Nevertheless we had
some new findings.  This is because we used least approximations
and took account of the vertical structure exactly.

Although our analysis is restricted in the axisymmetric mode,
the result will be helpful for more general analysis on the
stability of self-gravitating disk.  For later use we
summarize the main results in the following.

\begin{enumerate}
\item The Toomre-Hayashi model is unstable against axisymmetric
perturbation irrespectively of its model parameter, $ \gamma $.
\item The wavelength of the unstable perturbation is proportional
to the radius.  The ratio of the wavelength to the radius
is larger for smaller $ \gamma $.
\item Change in the vertical height can induce instability
even when the surface density does not change.
\item It is prospected that an unstable perturbation may grow
in proportion of time in the power of time.
\item The Toomre-Hayashi model of $ \gamma \, = \, 0 $,
i.e., the singular isothermal sphere is proven to be stable
against a non-spherical perturbation.
\end{enumerate}

The authors thank Tetsuro Konishi for telling us the asymptotic
solution shown in \S 3.
This research is financially supported
in part by the Grant-in-Aid for
Scientific Research on Priority Areas and that for 
Encouragement of Young Scientists of 
the Ministry of Education, Science, Sports and Culture of Japan
(Nos. 10147105, 11134209, 1202103, and 12740123).

\newpage

\begin{figure}
\epsscale{0.4}
\plotone{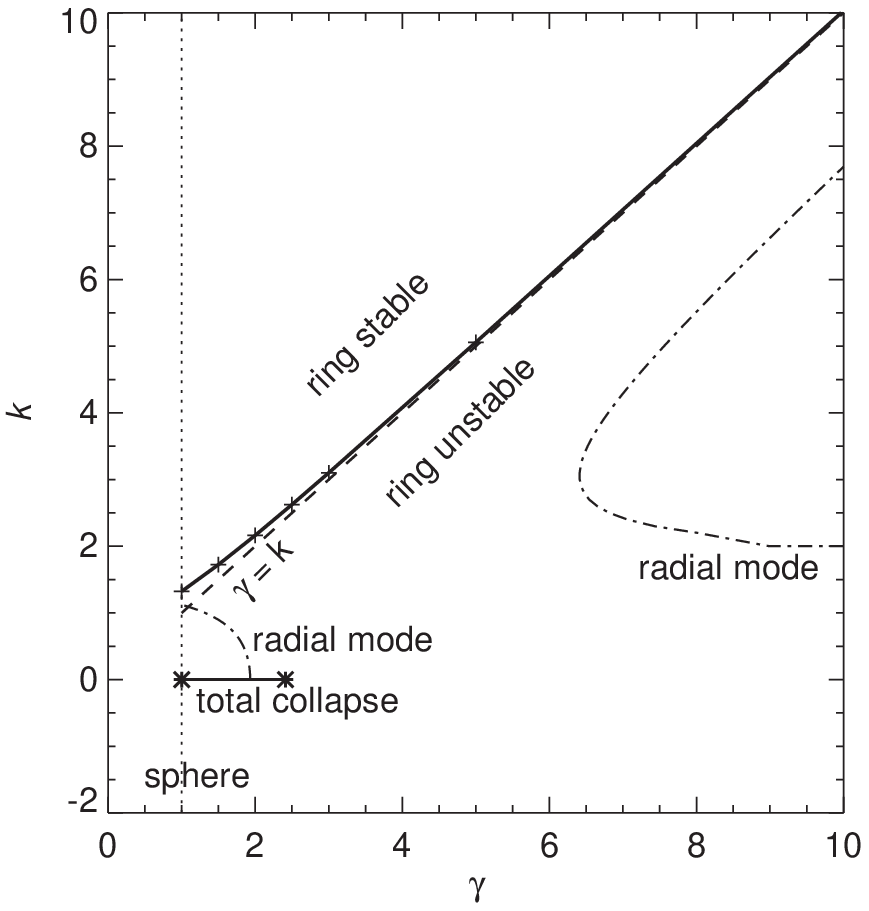} 
\caption{The critical wavenumber, $ k $, is shown as a
function of $ \gamma $.  The solid curve denotes
the wavenumber of the marginally stable mode.
The dashed curve denotes $ k \, = \, \gamma $ for
comparison.  The dash-dotted curves labeled
\lq radial mode' denotes the the marginally stable 
mode obtained by Shu et al. (2000) under the thin
disk approximation.}
\end{figure}

\begin{figure}
\epsscale{0.4}
\plotone{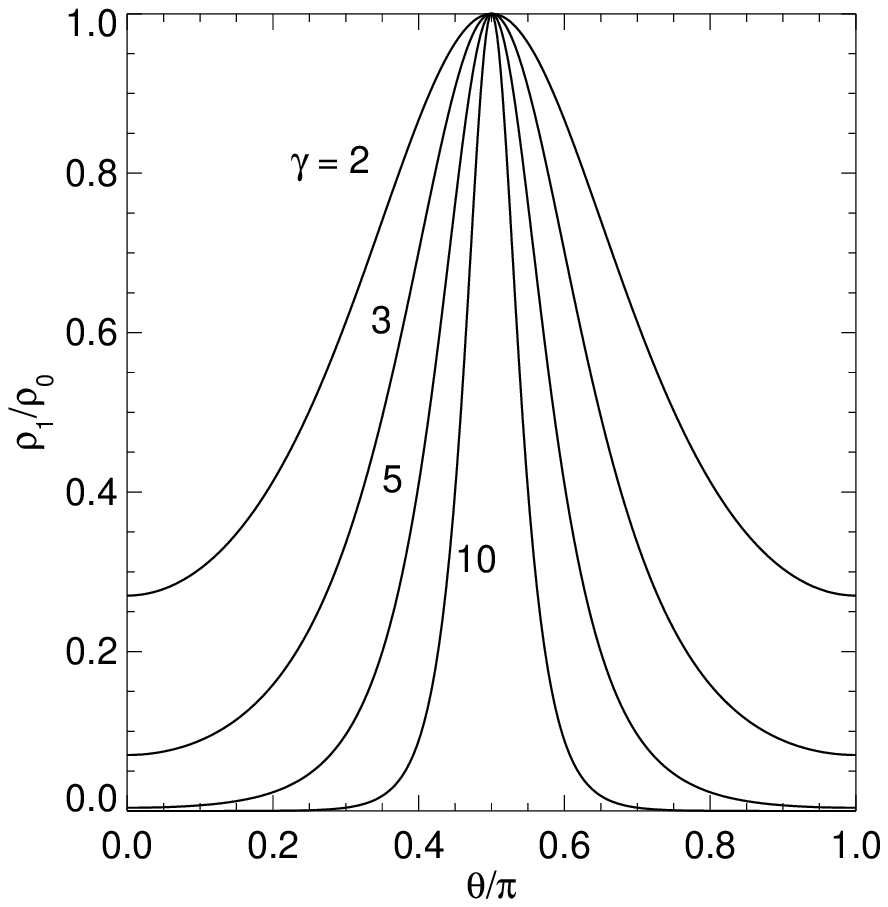}
\caption{The neutrally stable mode is shown for the Toomre-Hayashi
model of $ \gamma $ = 2, 3, 5, and 10.  The ordinate denotes
the relative density perturbation, $ \rho _1/\rho _0 $ while
the abscissa does $ \theta $.}  
\end{figure}

\begin{figure}
\epsscale{0.7}
\plotone{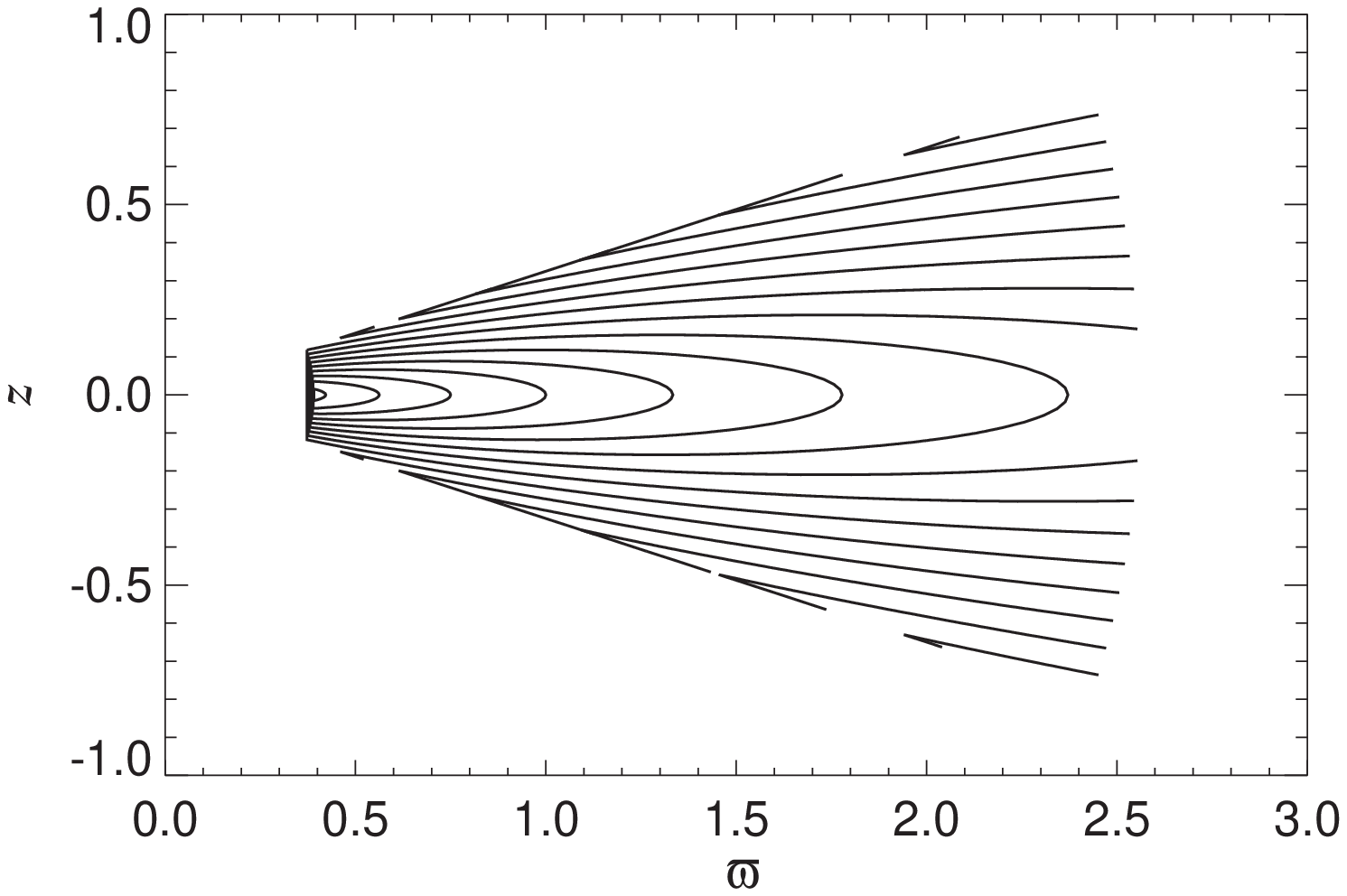} 
\caption{The density distribution in the $ \varpi \, - \, z $ plane
is shown for the Toomre-Hayashi model of $ \gamma \, = \, 10 $.
Note that this is drawn in the cylindrical coordinates.}
\end{figure}

\begin{figure}
\epsscale{0.7}
\plotone{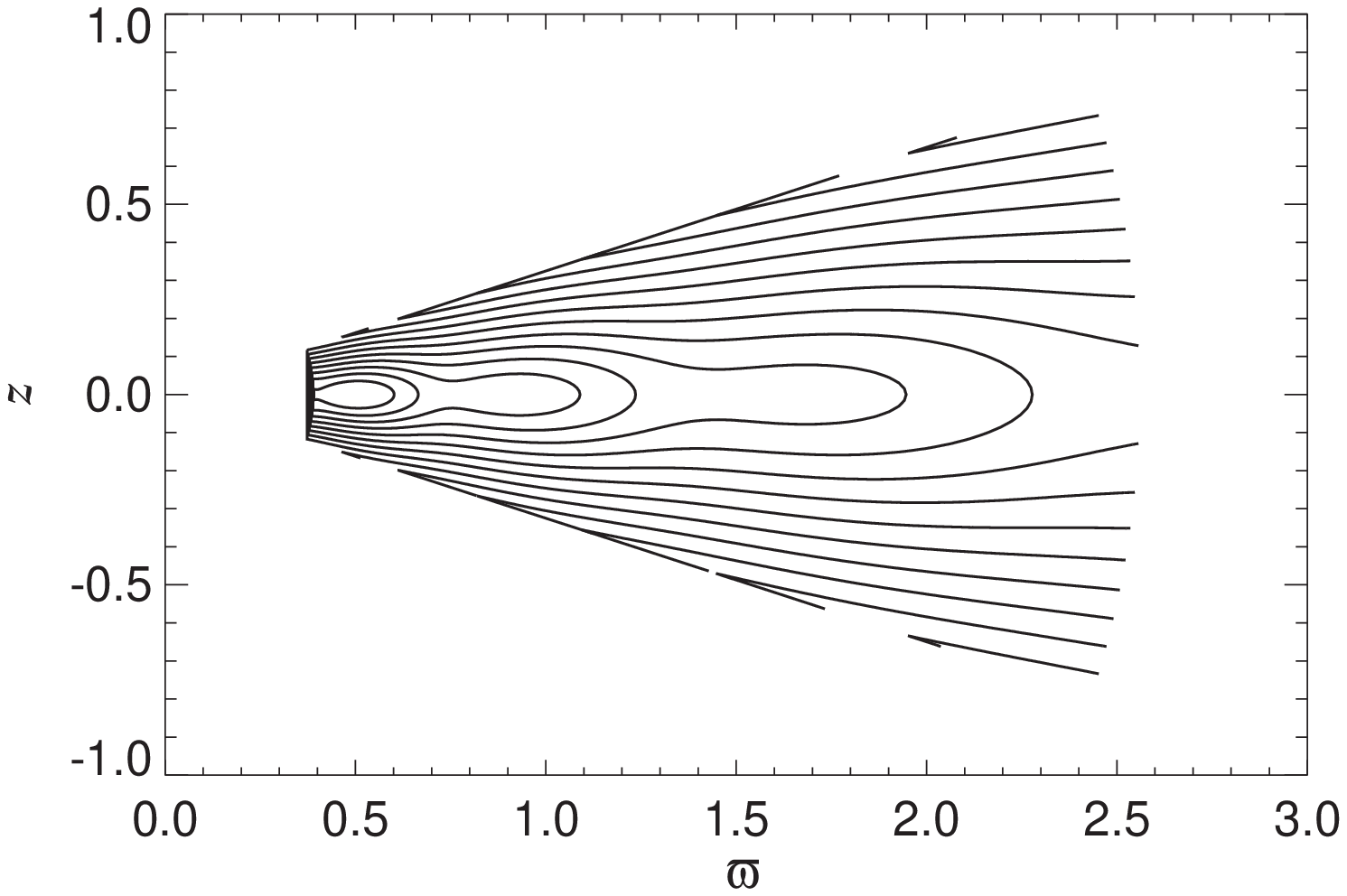} 
\caption{The same as Fig. 3 but the marginally stable mode is superimposed
on the equilibrium model.}
\end{figure}

\end{document}